\documentclass[USenglish,3p,onecolumn]{article}

\usepackage[utf8]{inputenc}
\usepackage{amsmath, amssymb}
\usepackage{amsfonts}
\usepackage{amsthm}
\usepackage{microtype}
\usepackage{graphicx}
\usepackage{color}

\newcommand{\mb}[1]{\mathbf{#1}}
\newcommand{\bs}[1]{\boldsymbol{#1}}
\newcommand{\vp}{\varphi_{\cP,\mb{s}}}
\newcommand{\cP}{\mathcal{P}}
\newcommand{\cF}{\mathcal{F}}
\newcommand{\cS}{\mathcal{S}}
\newcommand{\cT}{\mathcal{T}}
\newcommand{\bR}{\mathbb{R}}
\newcommand{\tp}{^{\mbox{\tiny\bf T}}}

\DeclareMathOperator*{\sgn}{sgn}
\DeclareMathOperator*{\vol}{Vol}

\newtheorem{lemma}{Lemma}[section]
\newtheorem{problem}{Problem}[section]
\newtheorem{corollary}{Corollary}[section]
\newtheorem{theorem}{Theorem}[section]
\newtheorem{remark}{Remark}[section]

\begin{document}

%
%
  
\title{Reconstruction of polytopes from the modulus of the Fourier transform with small wave length}

\author{Konrad Engel and Bastian Laasch\thanks{Universit\"at Rostock,  Institut f\"ur Mathematik, 18051 Rostock, Germany. E-mail: konrad.engel@uni-rostock.de and bastian.laasch@uni-rostock.de}}

\maketitle

  
\begin{abstract}
Let $\mathcal{P}$ be an $n$-dimensional convex polytope and $\mathcal{S}$ be a hypersurface in $\mathbb{R}^n$. This paper investigates potentials to reconstruct $\mathcal{P}$ or at least to compute significant properties of $\mathcal{P}$ if the modulus of the Fourier transform of $\mathcal{P}$ on $\mathcal{S}$ with wave length $\lambda$, i.e., $|\int_{\mathcal{P}} e^{-i\frac{1}{\lambda}\mathbf{s}\cdot\mathbf{x}} \,\mathbf{dx}|$ for $\mathbf{s}\in\mathcal{S}$, is given, $\lambda$ is sufficiently small and $\mathcal{P}$ and $\mathcal{S}$ have some well-defined properties. The main tool is an asymptotic formula for the Fourier transform  of $\mathcal{P}$ with wave length $\lambda$ when $\lambda \rightarrow 0$. The theory of X-ray scattering of nanoparticles motivates this study since the modulus of the Fourier transform of the reflected beam wave vectors are approximately measurable in experiments.
\end{abstract}
  
  %
%
%
%

\section{Introduction}

Let $\cP$ be an $n$-dimensional convex polytope in $\bR^n$. The \emph{Fourier transform of $\cP$} is defined by
\begin{align*}
    F_{\cP}(\mb{s})=\int_{\cP} e^{-i\mb{s}\cdot\mb{x}} \,\mb{dx}\,.
\end{align*}
Here the product $\cdot$ is the standard scalar product.
Moreover, the \emph{Fourier transform of $\cP$ with wave length $\lambda$} is the function
\begin{align*}
    \vp(\lambda)=F_{\cP}\left(\frac{1}{\lambda}\mb{s}\right)\,.
\end{align*}
It is well-known and obvious that $\vp(\lambda)$ tends to the volume of $P$ if $\lambda \rightarrow \infty$.

But in this paper we study the limit process $\lambda \rightarrow 0$ together with the following problem:

\begin{problem}[Reconstruction problem]\label{reconstruction_problem}
Assume that $|\vp(\lambda)|$ is given for a known fixed small positive $\lambda$ and for vectors $\mb{s}$ of a known proper subset $\cS$ of $\bR^n$, but for an unknown polytope $\cP$.
Determine the polytope $\cP$ or at least significant properties of $\cP$.
\end{problem}

The motivation to investigate Problem \ref{reconstruction_problem} is given by a physical application. In small-angle and also partially in wide-angle X-ray scattering of nanoparticles the modulus of the Fourier transform of the reflected beam wave vectors can be a ``approximately'' measured on the \emph{Ewald (half-)sphere} (see e.g. \cite{Raines2010}, \cite{Barke2015} and \cite{Rossbach2019}). Now the question arises whether conclusions can be drawn about the underlying particle based on its scattering pattern. In physics the introduced parameter $\lambda$ can be interpreted as \emph{wave length}, i.e., $\lambda\rightarrow 0$ implies frequency $\rightarrow\infty$. But note that in experiments $\lambda$ cannot be chosen arbitrarily small like in this paper (see \cite{Seibert2011}). 

There is already a vast amount of literature in this field. In \cite{Engel2020} it was shown that a $3$-dimensional convex polytope is uniquely determined by its scattering pattern on an arbitrarily small part of the Ewald sphere up to translation and reflection in a point, i.e., we just cannot distinguish between two polytopes $\cP$ and $\cP'$ for which there is a vector $\mb{v}$ and an $\epsilon \in \{-1,1\}$ so that
\begin{align*}
    \cP'=\epsilon \cP + \mb{v}\,.
\end{align*}

In \cite{Beinert2017} the reconstruction of a $1$-dimensional signal given the modulus of its Fourier transform was studied using the \emph{Prony Method}. Furthermore, the authors of \cite{Wischerhoff2016} developed an algorithm, also based on the Prony Method, to reconstruct $2$-dimensional non-convex polygons using the complex valued Fourier transform (not only the absolute value). Methods for $n$-dimensional convex polytopes and generalized polytopes given finitely many complex valued integral moments are presented in \cite{Gravin2012, Gravin2018}.

The Prony-type methods require that, for some specified lines through the origin, the value of the Fourier transform is known on sufficiently many points of the lines. But lines intersect spheres in at most two points and hence these methods cannot be applied if the (absolute) value of the Fourier transform is only known on a sphere.

A Machine Learning based method to reconstruct icosahedra from scattering data is described in \cite{Stielow2020}.

For our investigations, we first need some definitions. We say that a proper subset $\cS$ of $\bR^n$ is \emph{complete} if for every nonzero vector $\mb{s}$ in $\bR^n$ there is some $\mb{s}' \in \cS$ such that $\mb{s}$ is a scalar multiple of $\mb{s}'$. For example, a sphere around the origin is complete. Furthermore, we call an $n$-dimensional convex polytope \emph{facet-generic} if it does not contain two parallel facets, i.e., it does not contain two parallel $(n-1)$-dimensional faces. In this paper we solve the Reconstruction problem \ref{reconstruction_problem} ``approximately'' for facet-generic convex polytopes under the assumption that $\cS$ is complete. 

Let $\cF_{\cP}$ be the set of all facets of $\cP$. For a facet $F$ of $\cP$  let $\mb{p}_{F}$ be an arbitrary, but fixed point of the hyperplane containing $F$. If $\mb{s}$ is orthogonal to $F$ we denote this by $\mb{s} \bot F$. If $\mb{s} \bot F$ and $\mb{p}_{F}$ as well as $\mb{p}_{F}'$ are points of the hyperplane containing $F$, then $\mb{s} \cdot \mb{p}_{F} = \mb{s} \cdot \mb{p}_{F}'$ since $\mb{s} \cdot \left(\mb{p}_{F} - \mb{p}_{F}'\right) = 0$, and therefore we have the freedom to choose $\mb{p}_{F}$ arbitrarily on the hyperplane.

Let $\vol(\cP)$ be the volume of the convex polytope $\cP$ and let $A_{F}$ be the (positive) surface measure of its facet $F$. Moreover, we set 
\begin{align*}
    \sgn\,_{\cP}(F,\mb{s}) =
    \begin{cases}
        1&\text{if } \mb{s} \text{ is directed to the outside of }\cP\,,\\
        -1&\text{otherwise}\,.
    \end{cases} 
\end{align*}

In the following we do not explicitly write $\lambda \rightarrow 0$, because all limit processes in the paper are given in that way. The key result of the paper is the following:
\begin{theorem}\label{main_result}
Let $\cP$ be an $n$-dimensional convex polytope in $\bR^n$ and let $\mb{s} \neq \mb{0}$. Then
\begin{align*}
    \vp(\lambda)=\frac{i}{\|\mb{s}\|}\left(\sum_{F \in \cF_{\cP}: \mb{s} \bot F} \sgn\,_{\cP}(F,\mb{s}) A_F e^{-i\frac{1}{\lambda}\mb{s} \cdot \mb{p_F}}\right) \lambda + O(\lambda^2)\,.
\end{align*}
\end{theorem}

Note that the first item is an empty sum and hence vanishes if $\mb{s}$ is not orthogonal to any facet of $\cP$. Moreover, if $\cP$ is facet-generic, then the first item is either an empty sum or contains only one summand. This immediately implies:
\begin{corollary}\label{main_corollary}
Let $\cP$ be an $n$-dimensional facet-generic convex polytope in $\bR^n$ and let $\mb{s} \neq \mb{0}$. Then
\begin{align*}
    |\vp(\lambda)|=
    \begin{cases}
        \frac{A_F}{\|\mb{s}\|} \lambda + O(\lambda^2)&\text{if } \mb{s} \text{ is orthogonal to some facet } F\,,\\
        O(\lambda^2)&\text{otherwise}\,.
    \end{cases}
\end{align*}
\end{corollary}

The next section is devoted to the proof of Theorem \ref{main_result}. There we also discuss a generalisation to polytopal complexes. In Section 3 we study the reconstruction problem in more detail applying Corollary \ref{main_corollary}.

\section{Proof of Theorem \ref{main_result}}

Let
\begin{align*}
    I_{n,c}(\lambda)= \int_0^1 (1-x)^n e^{-i\frac{1}{\lambda} c x}\,dx\,.
\end{align*}

\begin{lemma}\label{integral_lemma}
Let $c$ be fixed nonzero real number and let $n$ be an integer with $n \ge 1$. Then
\begin{align*}
    I_{n,c}=-\frac{i}{c}\lambda + O(\lambda^2)\,.
\end{align*}
\end{lemma}

\proof
First note that by partial integration for $n \ge 1$
\begin{equation}\label{recursion}
    I_{n,c}(\lambda) = -\frac{i}{c}\lambda\left(1 - n I_{n-1,c}(\lambda)\right)
\end{equation}
and that 
\begin{align*}
    I_{0,c}(\lambda)=\int_0^1  e^{-i\frac{1}{\lambda} c x}\,dx = \frac{i}{c} \lambda \left(e^{-i \frac{1}{\lambda} c}-1\right)=O(\lambda)\,.
\end{align*}
Now we proceed by induction on $n$. If $n=1$, then we have by (\ref{recursion})
\begin{align*}
    I_{1,c}(\lambda) = -\frac{i}{c}\lambda\left(1 - I_{0,c}(\lambda)\right) = -\frac{i}{c}\lambda + O(\lambda^2)\,.
\end{align*}
The step from $n-1$ to $n$ follows analogously from (\ref{recursion}).\qed

For any $\mb{x} \in \bR^n$ let $\mb{x}'$ and $\mb{x}''$ be the vectors which can be obtained from $\mb{x}$ by deleting the last component and the last two components, respectively.

First we prove the assertion for the unit simplex $\triangle_n=\{\mb{x}\in \bR^n: \mb{1} \cdot \mb{x}\le 1$ and $ \mb{x} \ge \mb{0}\}$. Let $\mb{e}_j$ be the $j$-th unit vector, $j \in [n]$.

{\bf Case 1} The vector $\mb{s}$ is orthogonal to some facet. Without loss of generality we may assume that $\mb{s}$ is directed to the outside of $\triangle_n$.
Then there is some $\mu > 0$ such that $\mb{s} = -\mu \mb{e}_j$ for some $j \in [n]$ or $\mb{s}=\mu \mb{1}$.

{\bf Case 1.1} (Without loss of generality) $\mb{s} = -\mu \mb{e}_n$. Then $F=\{\mb{x} \in \mb{R}^n: \mb{x}' \in \triangle_{n-1} \text{ and } x_n=0\}$ is the corresponding facet and
$A_F=\vol(\triangle_{n-1})$. Moreover we may choose $\mb{p}_F=\mb{0}$ and hence $e^{-i\frac{1}{\lambda}\mb{s} \cdot \mb{p_F}}=1$. We have by iterated integration and Lemma \ref{integral_lemma}
\begin{align}
    \nonumber
    \vp(\lambda) &=\int_0^1\left(\int_{(1-x_n) \triangle_{n-1}} e^{-i \frac{1}{\lambda}(-\mu) x_n} \,\mb{dx}'\right) \,dx_n\\
    \nonumber    &=\vol(\triangle_{n-1})\int_0^1 (1-x_n)^{n-1}e^{-i \frac{1}{\lambda}(-\mu) x_n}  \,dx_n\\
    \nonumber    &=\vol(\triangle_{n-1}) I_{n-1,-\mu}(\lambda)\\
    \label{1.1}  &=\frac{i}{\mu} \vol(\triangle_{n-1}) \lambda + O(\lambda^2)\\
    \nonumber    &=\frac{i}{\|\mb{s}\|} A_F e^{-i\frac{1}{\lambda}\mb{s} \cdot \mb{p_F}}\lambda + O(\lambda^2)\,,
\end{align}
as desired.

{\bf Case 1.2} $\mb{s}=\mu \mb{1}$. Then $F=\{\mb{x} \in \mb{R}^n: \mb{x} \ge \mb{0} \text{ and }x_1+\dots+x_n=1\}$ is the corresponding facet. Moreover, we may choose
$\mb{p}_F=\frac{1}{n}\mb{1}$ and hence $e^{-i\frac{1}{\lambda}\mb{s} \cdot \mb{p_F}} = e^{-i\frac{1}{\lambda} \mu}$. With the transformation $y_i=x_i$, $i \in [n-1]$, and $y_n=1-(x_1+\dots+x_n)$ we obtain analogously as before
\begin{align*}
    \vp(\lambda)&=\int_0^1\left(\int_{(1-y_n) \triangle_{n-1}} e^{-i \frac{1}{\lambda}\mu y_n} \,\mb{dy}'\right) \,dy_n\\
    &=\vol(\triangle_{n-1})\int_0^1 (1-y_n)^{n-1}e^{-i \frac{1}{\lambda}\mu (1-y_n)}  \,dy_n\\
    &=\vol(\triangle_{n-1})e^{-i \frac{1}{\lambda}\mu}I_{n-1,-\mu}(\lambda)\\
    &=\frac{i}{\mu} \vol(\triangle_{n-1}) e^{-i \frac{1}{\lambda}\mu}\lambda + O(\lambda^2)\\
    &=\frac{i}{\|\mb{s}\|} A_F e^{-i\frac{1}{\lambda}\mb{s} \cdot \mb{p_F}}\lambda + O(\lambda^2)\,.
\end{align*}

{\bf Case 2} The vector $\mb{s}$ is not orthogonal to any facet. Let without loss of generality $s_n \neq 0$. By iterated integration
\begin{align*}
    \vp(\lambda)&=\int_{\triangle_{n-1}} \left(\int_{0}^{1-x_1-\dots-x_{n-1}} e^{-i \frac{1}{\lambda} (\mb{s}'\cdot \mb{x}'+s_nx_n)}\,dx_n\right) \,\mb{dx}'\\
    &=\frac{i}{s_n} \lambda \int_{\triangle_{n-1}} \left(e^{-i \frac{1}{\lambda} ((\mb{s}'-s_n \mb{1}')\cdot \mb{x}'+s_n)} - e^{-i \frac{1}{\lambda} \mb{s}'\cdot \mb{x}'}\right) \,\mb{dx}'\,.
\end{align*}

Now we study the integration of both items separately. We start with the first item.
We have $\mb{s}'-s_n \mb{1}' \neq \mb{0}'$ because otherwise $\mb{s}=s_n \mb{1}$ in contradiction to the assumption that $\mb{s}$ is not orthogonal to a facet.
Let without loss of generality $s_{n-1}-s_n \neq 0$. Again by iterated integration
\begin{align*}
    &\int_{\triangle_{n-1}} e^{-i \frac{1}{\lambda} ((\mb{s}'-s_n \mb{1}')\cdot \mb{x}'+s_n)} \mb{dx}' \\
    &=e^{-i \frac{1}{\lambda} s_n}\int_{\triangle_{n-2}} \left(\int_0^{1-x_1-\dots-x_{n-2}}e^{-i \frac{1}{\lambda} ((\mb{s}''-s_n \mb{1}'')\cdot \mb{x}''+(s_{n-1}-s_n)x_{n-1})}\,dx_{n-1} \right)\, \mb{dx}'' \\
    &=\frac{i}{s_{n-1}-s_n} \lambda e^{-i \frac{1}{\lambda} s_n}\int_{\triangle_{n-2}} e^{-i \frac{1}{\lambda} ((\mb{s}''-s_n \mb{1}'')\cdot \mb{x}'')} 
    (e^{-i \frac{1}{\lambda} (s_{n-1}-s_n)(1-x_1-\dots-x_{n-2})}-1)\, \mb{dx}''\\
    &=O(\lambda)\,.
\end{align*}

We treat the second item in an analogous way. By the assumption $\mb{s}$ is not a multiple of a unit vector. Thus we may assume without loss of generality that $s_{n-1} \neq 0$. Then 
\begin{align*}
    &\int_{\triangle_{n-1}} e^{-i \frac{1}{\lambda} \mb{s}'\cdot \mb{x}'}\mb{dx}' \\
    &=\int_{\triangle_{n-2}} \left(\int_0^{1-x_1-\dots-x_{n-2}}e^{-i \frac{1}{\lambda} (\mb{s}''\cdot \mb{x}''+s_{n-1}x_{n-1})}\,dx_{n-1} \right)\, \mb{dx}'' \\
    &=\frac{i}{s_{n-1}} \lambda \int_{\triangle_{n-2}} e^{-i \frac{1}{\lambda} (\mb{s}''\cdot \mb{x}'')} 
    (e^{-i \frac{1}{\lambda} s_{n-1}(1-x_1-\dots-x_{n-2})}-1)\, \mb{dx}''\\
    &=O(\lambda)\,.
\end{align*}

Consequently
\begin{align*}
    \vp(\lambda)=O(\lambda^2)\,,
\end{align*}
as desired. 

Now we prove the assertion for an arbitrary simplex $\cP$ with vertices $\mb{v}_0,\dots,\mb{v}_n$.

{\bf Case 1} The vector $\mb{s}$ is orthogonal to some facet $F$ of $\cP$. Without loss of generality we may assume that $F$ is spanned by $\mb{v}_0,\dots,\mb{v}_{n-1}$ and that $\mb{s}$ is directed to the outside of $\triangle_n$. We choose $\mb{p}_F=\mb{v}_0$. Let $d$ be the distance between $\mb{v}_n$ and $F$. 
Using the Hessian normal form we obtain
\begin{align*}
    d=-\frac{(\mb{v}_n-\mb{v}_0) \cdot \mb{s}}{\|\mb{s}\|}\,.
\end{align*}
Let $T$ be the matrix whose $j$-th column is $\mb{v}_j - \mb{v}_0$, $j \in [n]$.
Then
\begin{align*}
    \frac{1}{n!}|\det(T)|=\vol(\cP) = \frac{1}{n} A_F d
\end{align*}
and consequently
\begin{align}\label{A_F}
    A_F=-\frac{1}{(n-1)!} |\det(T)| \frac{\|\mb{s}\|}{(\mb{v}_n-\mb{v}_0) \cdot \mb{s}}\,.
\end{align}
The affine transformation
\begin{align*}
    \mb{x} = T \mb{y} + \mb{v}_0
\end{align*}
maps $\triangle_n$ onto $\cP$ and we have
\begin{align*}
    \int_{\cP} e^{-i\frac{1}{\lambda}\mb{s}\cdot\mb{x}} \,\mb{dx}
    &=|\det(T)|\int_{\triangle_n} e^{-i\frac{1}{\lambda}\mb{s}\cdot (T \mb{y} + \mb{v}_0)} \,\mb{dy}\\
    &=|\det(T)|e^{-i\frac{1}{\lambda}\mb{s}\cdot\mb{p}_F}\int_{\triangle_n} e^{-i\frac{1}{\lambda}(T\tp\mb{s})\cdot \mb{y}} \,\mb{dy}\,,
\end{align*}
i.e.,
\begin{equation}\label{vp}
    \vp(\lambda)=|\det(T)|e^{-i\frac{1}{\lambda}\mb{s}\cdot\mb{p}_F}\varphi_{\triangle_n, T\tp \mb{s}}(\lambda).
\end{equation}
Since $\mb{s}$ is orthogonal to $F$, i.e., to $\mb{v}_j-\mb{v}_0$, $j \in [n-1]$, we have 
\begin{align*}
    T\tp\mb{s}=\left((\mb{v}_n-\mb{v}_0)\cdot \mb{s}\right)\mb{e}_n\,.
\end{align*}
From Case 1.1  for the  unit simplex (see (\ref{1.1})) we know that
\begin{align}\label{vp_triangle}
    \varphi_{\triangle_n, T\tp \mb{s}}(\lambda)=\frac{i}{-(\mb{v}_n-\mb{v}_0)\cdot \mb{s}} \frac{1}{(n-1)!} \lambda + O(\lambda^2)\,.
\end{align}
Thus (\ref{A_F})--(\ref{vp_triangle}) imply
\begin{align*}
    \vp(\lambda)=\frac{i}{\|\mb{s}\|} A_F e^{-i\frac{1}{\lambda}\mb{s} \cdot \mb{p_F}}\lambda + O(\lambda^2)\,.
\end{align*}

{\bf Case 2}
The vector $\mb{s}$ is not orthogonal to any facet of $\cP$. We may argue in the same way as for Case 1. We only have to verify that $T\tp \mb{s}$ is not orthogonal to any facet of $\triangle_n$.

Assume that $T\tp \mb{s}$ is a scalar multiple of $\mb{e}_k$ for some $k \in [n]$. Then $(\mb{v}_j-\mb{v}_0) \cdot \mb{s} = 0$ for all $j \in [n] \setminus \{k\}$, and hence $\mb{s}$ is orthogonal to the facet spanned by the vertices $\mb{v}_j$ with $j \in \{0,\dots,n\}\setminus \{k\}$, a contradiction.

Now, assume that $T\tp \mb{s}$ is a scalar multiple of $\mb{1}$. Then $(\mb{v}_1-\mb{v}_0) \cdot \mb{s} = \cdots = (\mb{v}_n-\mb{v}_0) \cdot \mb{s}$ which implies $(\mb{v}_j-\mb{v}_1) \cdot \mb{s}=0$ for all $j \in \{2,\dots,n\}$. Consequently
$\mb{s}$ is orthogonal to the facet spanned by the vertices $\mb{v}_j$ with $j \in [n]$, a contradiction.

Finally we prove the assertion for an arbitrary convex polytope $\cP$. This can be easily done using a triangulation of $\cP$, i.e., a set $\Theta$ of simplices with the following properties: The union of all members is $\cP$ and any two members are either disjoint or intersect in a common face. Let $\cF_\Theta$ be the set of all facets of the simplices in the triangulation. Using the proved assertion for simplices we obtain
\begin{align*}
    \vp(\lambda) &= \sum_{\cT \in \Theta} \varphi_{\cT,\mb{s}}(\lambda)\\
    &=\frac{i}{\|\mb{s}\|}\sum_{\cT \in \Theta}\left(\sum_{F \in \cF_{\cT}: \mb{s} \bot F} \sgn\,_{\cT}(F,\mb{s}) A_F e^{-i\frac{1}{\lambda}\mb{s} \cdot \mb{p_F}}\right) \lambda + O(\lambda^2)\\
    &=\frac{i}{\|\mb{s}\|}\sum_{F \in \cF_\Theta: \mb{s} \bot F}A_F e^{-i\frac{1}{\lambda}\mb{s} \cdot \mb{p_F}}\left(\sum_{\cT \in \Theta: F \in \cF_{\cT}} \sgn\,_{\cT}(F,\mb{s}) \right) \lambda + O(\lambda^2)\,.
\end{align*}

We say that $F \in \cF_{\Theta}$ is \emph{visible} if $F$ is the facet of only one simplex and it is \emph{invisible} if it is a facet of exactly two simplices. Obviously, each $F \in \cF_{\Theta}$ is either visible or invisible. Thus the inner sum contains only one or two items.

Let $F \in \cF_{\Theta}$ be an invisible facet and concretely a facet of the simplices $\cT_1$ and $\cT_2$ of the triangulation. If $\mb{s} \bot F$, then obviously $\sgn\,_{\cT_1}(F,\mb{s})=-\sgn\,_{\cT_2}(F,\mb{s})$. Thus the inner sum vanishes. This means that the contribution of invisible facets is only of order $O(\lambda^2)$.

Let $\check{\cF}_{\Theta}$ be the set of all visible facets. For each $F' \in \check{\cF}_{\Theta}$ let $\cT_{F'}$ be the unique simplex of the triangulation having $F'$ as facet. Each $F' \in \check{\cF}_{\Theta}$ is part of exactly one facet $F$ of $P$ and for each facet $F$ of $\cP$
\begin{align*}
    A_F=\sum_{F' \in \check{\cF}_{\Theta}: F' \subseteq F} A_{F'}\,.
\end{align*}

Note that we may choose $\mb{p}_{F'} = \mb{p}_F$ if $F' \subseteq F$. Moreover, if
$\mb{s} \bot F$ and  $F' \subseteq F$, then $\sgn\,_{\cT_{F'}}(F',\mb{s})=\sgn\,_{\cP}(F,\mb{s})$. Thus we may continue the computation of $\vp(\lambda)$ and obtain
\begin{align*}
    \vp(\lambda) &=\frac{i}{\|\mb{s}\|}\left(\sum_{F' \in \check{\cF}_\Theta: \mb{s} \bot F'}A_{F'} e^{-i\frac{1}{\lambda}\mb{s} \cdot \mb{p_{F'}}} \sgn\,_{\cT_{F'}}(F',\mb{s})\right) \lambda + O(\lambda^2)\\
    &=\frac{i}{\|\mb{s}\|}\left(\sum_{F \in \cF_{\cP}: \mb{s} \bot F}\sum_{F' \in \check{\cF}_\Theta: F' \subseteq F}A_{F'} e^{-i\frac{1}{\lambda}\mb{s} \cdot \mb{p_{F'}}} \sgn\,_{\cT_{F'}}(F',\mb{s})\right) \lambda + O(\lambda^2)\\
    &=\frac{i}{\|\mb{s}\|}\left(\sum_{F \in \cF_{\cP}: \mb{s} \bot F}e^{-i\frac{1}{\lambda}\mb{s} \cdot \mb{p_{F}}} \sgn\,_{\cP}(F,\mb{s})\sum_{F' \in \check{\cF}_\Theta: F' \subseteq F}A_{F'} \right) \lambda + O(\lambda^2)\\
    &=\frac{i}{\|\mb{s}\|}\left(\sum_{F \in \cF_{\cP}: \mb{s} \bot F}\sgn\,_{\cP}(F,\mb{s}) A_F e^{-i\frac{1}{\lambda}\mb{s} \cdot \mb{p_{F}}}  \right) \lambda + O(\lambda^2)\,.
\end{align*}
Thus the whole proof is completed. \qed

The result can easily be generalised to polytopal complexes: Here we consider a \emph{polytopal complex} as a finite union of $n$-dimensional convex polytopes such that any two of them are either disjoint or intersect in a common face (polytopes of smaller dimension do not contribute to the integral). As for triangulations we may define visible and invisible facets. With the same arguments as for triangulations of convex polytopes we may derive that the contribution of invisible facets is only of order $O(\lambda^2)$. Thus in Theorem \ref{main_result} $\cF_{\cP}$ can be replaced by $\check{\cF}_{\cP}$, i.e., the set of all visible facets of the polytopal complex $\cP$.

\section{``Approximative'' solution of the reconstruction problem for facet-generic convex polytopes}

In the title we write \emph{approximative} between quotation marks because we study this problem more from a practical point of view. A detailed estimation of the error terms remains an open problem for the future.

We restrict ourselves to \emph{minimal} complete subsets $\cS$ of $\bR^n$, i.e.,  for every nonzero vector $\mb{s}$ in $\bR^n$ there is exactly one $\mb{s}' \in \cS$ such that $\mb{s}$ is a scalar multiple of $\mb{s}'$. For example, a hemisphere around the origin (and having only a ``half'' from the equator) is minimal complete. Moreover, a sphere touching a coordinate hyperplane at the origin without the origin is almost minimal complete. This concerns in particular the Ewald sphere. Here we must write \emph{almost} because we do not have a corresponding $\mb{s}'$ for the vectors $\mb{s}$ lying in this coordinate hyperplane.

For practical reasons we assume that $\cS$ is parameterisable, i.e., we can write $\cS$ in the form
\begin{align*}
    \cS=\{\bs{\sigma}(\mb{t}): \mb{t} \in D\}\,,   
\end{align*}

where $D \subseteq \bR^{n-1}$ is the domain of the vector function $\bs{\sigma}: \bR^{n-1} \rightarrow \bR^n$. Finally we assume that
$\bs{\sigma}: D \rightarrow \cS$ is bijective.

\subsection{Computation of significant information of a facet-generic convex polytope based on the modulus of its Fourier transform}\label{significant_information_section}

Now let $\cP$ be an unknown facet-generic convex polytope with (unknown) $f$ facets and assume that we know for some fixed small wave length $\lambda$ the values of $|\vp(\lambda)|$ for all $\mb{s} \in \cS$ (resp. for all $\mb{s}$ in a sufficiently ``dense'' finite subset of $\cS$). Then we know also the values of
\begin{align}\label{psi_value}
    \bs{\psi}_{\lambda}(\mb{t})=\frac{1}{\lambda} \|\bs{\sigma}(\mb{t})\||\varphi_{\cP,\bs{\sigma}(\mb{t})}(\lambda)|
\end{align}
for all $\mb{t} \in D$ (resp. for all $\mb{t}$ in a sufficiently ``dense'' finite subset of $D$).

From Corollary \ref{main_corollary} we obtain that
\begin{align*}
    \lim_{\lambda \rightarrow 0}\bs{\psi}_{\lambda}(\mb{t})=
    \begin{cases}
    A_F&\text{if } \bs{\sigma}(\mb{t}) \text{ is orthogonal to some facet } F\,,\\
    0&\text{otherwise}\,.
    \end{cases} 
\end{align*}

Thus we can expect that, also for our small positive wave length $\lambda$, we can find exactly $f$ significant maximum points $\mb{t}_j$, $j \in[f]$, of $\bs{\psi}_{\lambda}(\mb{t})$. These points can be found with two methods using a well chosen bound $\theta$, where $\theta$ is positive, but smaller than the expected minimal surface area of a facet of the unknown polytope:
\begin{itemize}
    \item We sufficiently smooth the function $\bs{\psi}_{\lambda}(\mb{t})$ and determine all local maximum points $\mb{t}_j$ with $\bs{\psi}_{\lambda}(\mb{t}_j) > \theta$.
    \item We determine all local maximum points $\mb{t}$ with $\bs{\psi}_{\lambda}(\mb{t}) > \theta$ and cluster them in such a way that each cluster $C_j$ contains points with small mutual distance. For each cluster $C_j$ we choose a point $\mb{t}_j$ for which $\bs{\psi}_{\lambda}(\mb{t}_j) \ge \bs{\psi}_{\lambda}(\mb{t})$ for all $\mb{t} \in C_j$.
\end{itemize}

Then the vector $\bs{\sigma}(\mb{t}_j)$ is almost orthogonal to the (unknown) facet $F_j$, $j \in [f]$. Moreover, $A_j=\bs{\psi}_{\lambda}(\mb{t}_j)$ is an approximation of $A_{F_j}$. Of course, we can normalise $\bs{\sigma}(\mb{t}_j)$ and obtain a vector $\mb{n}_j$. Thus we have a set 
\begin{align*}
    I=\{(\mb{n}_j,A_j): j \in [f]\}
\end{align*}
which contains significant information on $\cP$, namely approximations $\mb{n}_j$ of the normal vectors and approximations of the corresponding facet areas $A_j$. Please note that $\mb{n}_j$ can be an outward or an inward normal vector of $\cP$. Hence, its ``right'' direction, i.e., its sign, is still unknown.

We say that a set $I$ is a \emph{facet-indicator set} of the facet-generic convex polytope $\cP$ if $\mb{n}_j$ is a unit normal vector (outwards \textbf{or} inwards) of a facet of $\cP$ and $A_j$ is the surface measure of this facet, $j \in [f]$.

\begin{figure}[h]
    \centering
    \includegraphics[height=4.45cm]{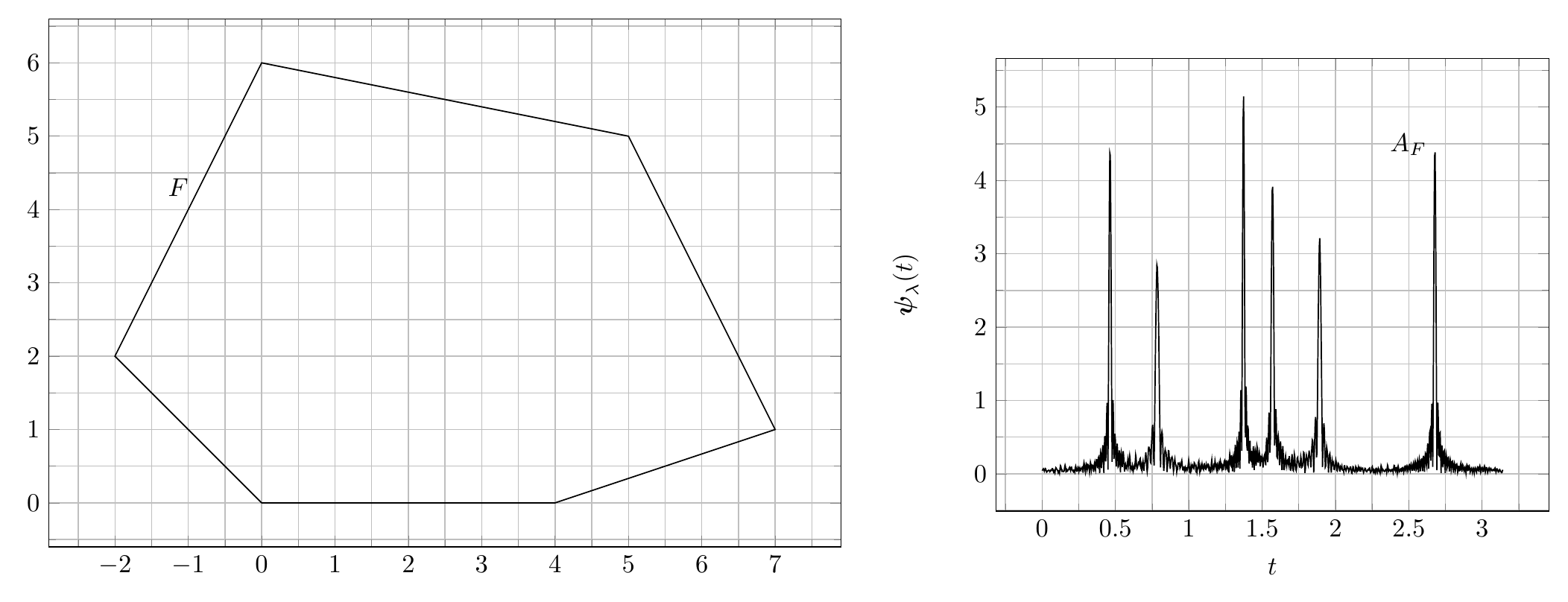}
    \caption{A $2$-dimensional facet-generic convex polygon with six facets (left) and the corresponding values of $\bs{\psi}_{\lambda}(t)$ given by (\ref{psi_value}) (right). The local maximum for the edge $F$ is marked by $A_F$.}
    \label{2D_example_for_psi}
\end{figure}

\textbf{Example} A numerical example for the $2$-dimensional case is given in Figure \ref{2D_example_for_psi}. For the given polygon with six sides (facets) the corresponding values of $\bs{\psi}_{\lambda}(\mb{t})$ were calculated. It can be seen that there are six local maxima, each for one side. The function $\bs{\sigma}: \bR \rightarrow \bR^2$ was chosen as
\begin{align*}
    \bs{\sigma}(t)= \begin{pmatrix} \cos(t)\\ \sin(t)\end{pmatrix}, t\in[0, \pi)\,.
\end{align*}
Hence, the set 
\begin{align*}
    \cS=\{\bs{\sigma}(t): t \in [0, \pi)\}\,,
\end{align*}
corresponds to a semicircle. The wave length parameter $\lambda$ was set to $0.01$.

For the computation of the Fourier transform in the 2- and 3-dimensional case we used the following formulas that can be obtained using integral theorems (see e.g. \cite{Wuttke2017}).
The computation of the Fourier transform of $n$-dimensional polytopes is presented e.g. in \cite{Beck2007} and \cite{Barvinok2008}.

\begin{remark}
Let $\cP$ be a polygon and $\mathcal{E}_{\cP}$ be the set of the positive oriented edges of $\cP$ (interpreted as vectors of $\bR^2$). For $\mb{e} \in \mathcal{E}_{\cP}$  let $\mb{e}^-$ be its starting point and  $\mb{e}^+$ be its end point. Let $\mb{s}^{\bot}=(-s_2, s_1)^T$ ($\mb{s}^{\bot}$ is orthogonal to $\mb{s}$).  Then
\begin{align*}
    F_{\cP}(\mb{s})=\int_{\cP} e^{-i\mb{s}\cdot\mb{x}} \,\mb{dx}=
		-\frac{\mb{s}^{\bot}}{\|\mb{s}\|^2}\cdot\sum_{\mb{e}\in\mathcal{E}_{\cP}} \mb{e}\left(\frac{e^{-i\mb{s}\cdot\mb{e}^+} - e^{-i\mb{s}\cdot\mb{e}^-}}{\mb{s}\cdot\mb{e}}\right)\,.
\end{align*}
Let $\cP$ be a $3$-dimensional convex polytope and $\cF_{\cP}$ the set of its facets. For $F\in\cF_{\cP}$ let  $\mb{n}_F$ be the outer normal vector of $\cP$ being orthogonal to $F$ and assume that the edges of $\mathcal{E}_F$ are positively oriented with respect to $\mb{n}_F$.  Then for all $F\in\cF_{\cP}$ the Fourier transform $F_{\cP}(\mb{s})$ is given by
\begin{align*}
    -i \frac{\mb{s}}{\|\mb{s}\|^2}\cdot\sum_{F\in\cF_{\cP}} \mb{n}_F \left(\frac{\left(\mb{n}_F\times\mb{s}\right)}{\|\mb{s}\|^2-(\mb{s}\cdot\mb{n}_F)^2}\cdot\sum_{\mb{e}\in\mathcal{E}_F}\mb{e}\left(\frac{e^{-i\mb{s}\cdot\mb{e}^+} - e^{-i\mb{s}\cdot\mb{e}^-}}{\mb{s}\cdot\mb{e}}\right)\right)\,.
\end{align*}
In both cases the right side has to be interpreted as a limit if some denominator vanishes. Moreover, the product $\cdot$ is here the dot product without conjugation of the second factor, i.e., $\mb{x} \cdot \mb{y} = \sum_{i=1}^n x_iy_i$ also if the second factor is complex.
\end{remark}

\subsection{Reconstruction of facet-generic convex polytopes}\label{reconstruction_section}

In the following we want to investigate reconstruction potentials based on a facet-indicator set $I$ of an unknown polytope $\cP$. Consequently, we are led to the following problem:
\begin{problem}\label{indicator_problem}
For a given set $I=\{(\mb{n}_j,A_j): j \in [f]\}$ determine a facet-generic convex polytope $\cP$ such that $I$ is a facet-indicator set of $\cP$. 
\end{problem}

\begin{figure}[h]
    \centering
    \includegraphics[height=4.45cm]{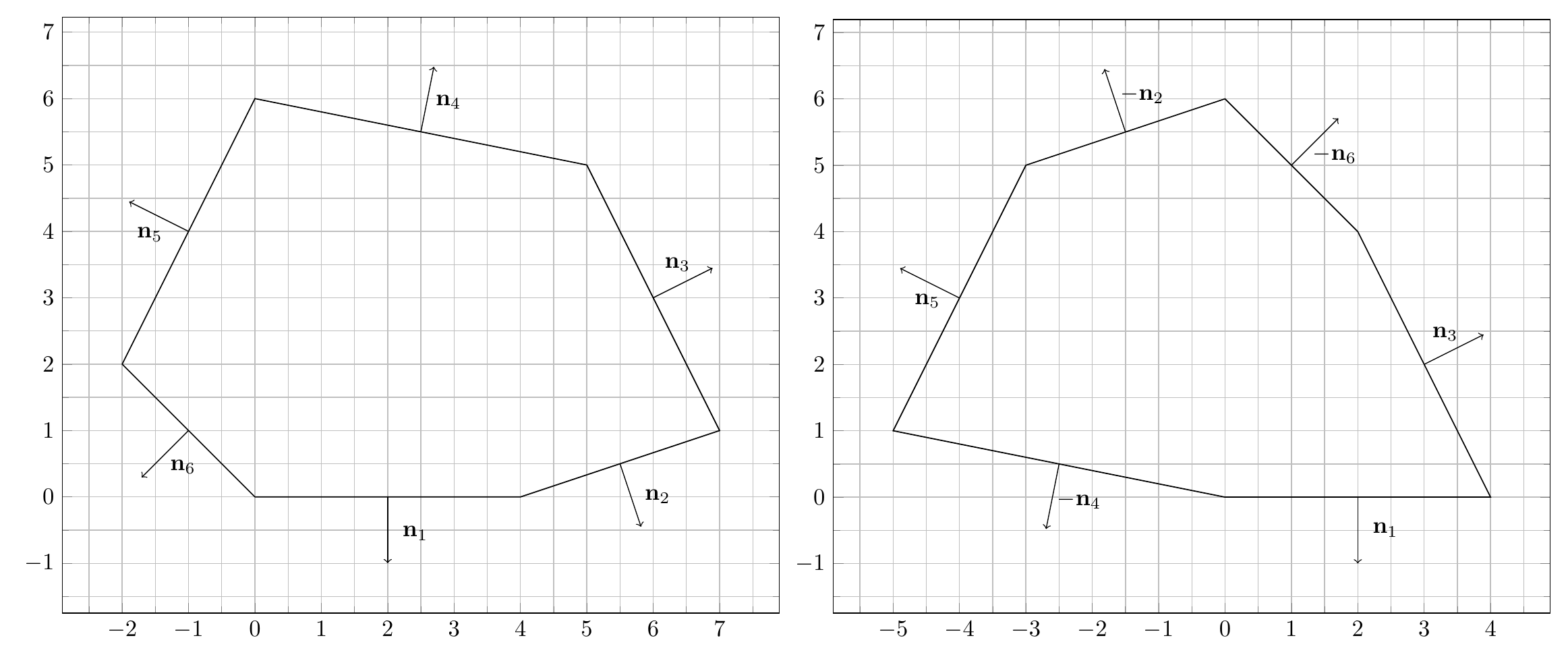}
    \caption{Two $2$-dimensional facet-generic convex polygons having -- up to the sign -- the same unit normal vectors and the same corresponding side lengths, i.e., they have the same facet-indicator set.}
    \label{2D_example_ambiguity}
\end{figure}

Since the set $I$ only gives us the normal vectors but not their outer directions, there are examples of ambiguities, i.e., a unique reconstruction is impossible. For a given $I$ there can be multiple facet-generic convex polytopes with different structures with $I$ as their facet-indicator set even with an exact calculation (see Figure \ref{2D_example_ambiguity}).

However, for $n$-dimensional simplices the following theorem proves a uniqueness result up to translation and reflection in a point. The proof also contains a solution of Problem \ref{indicator_problem}.

\begin{theorem}\label{simplex_reconstruction_theorem}
Let $I=\{(\mb{n}_j,A_j): j \in \{0,\dots,n\}\}$ be a facet-indicator set of an unknown $n$-dimensional simplex $\cP$ in $\bR^n$. Then $\cP$ can be uniquely determined up to translation and reflection in a point.
\end{theorem}

\proof We may assume without loss of generality that $\mb{0}$ is a vertex of $\cP$. Then $\cP$ is the feasible set of a system of the following form
\begin{eqnarray*}
    \mb{n}_0 \cdot \mb{x} & R_0& a,\\
    \mb{n}_1 \cdot \mb{x} & R_1& 0,\\
    \ldots&\\
    \mb{n}_n \cdot \mb{x} & R_n& 0\,,
\end{eqnarray*}
where $R_j \in \{\le, \ge\}$, $j \in \{0,\dots,n\}$. Since we may replace, if necessary, $\cP$ by $-\cP$ we may assume that $R_0$ is the relation $\le$. Moreover we have $a > 0$ because $\mb{0} \in \cP$.

Recall that we assume that $\mb{v}_0=\mb{0}$ is a vertex. Let the other (unknown) vertices of $\cP$ be $\mb{v}_1,\dots,\mb{v}_n$, where they are labeled in such a way that the vertices $\mb{v}_k$, $k \in \{0,\dots,j-1,j+1,\dots,n\}$, belong to the facet with normal vector $\mb{n}_j$. 
Then
\begin{align}\label{nv}
    \mb{n}_j \cdot \mb{v}_k=
    \begin{cases}
        a&\text{if }0=j \neq k\,,\\
        0&\text{if }1\le j \neq k\,.
    \end{cases}
\end{align}
If we are able to decide whether $\mb{n}_j \cdot \mb{v}_j$ is positive or negative, then we know that we have to take for $R_j$ the relation $\le$ or $\ge$, respectively, $j \in [n]$. If, in addition, we are able to determine $a$, then the system and hence also $\cP$ are uniquely determined. Even without computing the vertices explicitly this can be done as follows:

Let $N_j$ (resp. $V_j$) be the matrix whose columns are the vectors $\mb{n}_k$ (resp. $\mb{v}_k$), $k = 0,\dots,j-1,j+1,\dots,n$. Now fix some $j \in \{1,\dots,n\}$. Clearly, there is some unique vector $\bs{\lambda}_j$ such that
\begin{align}\label{lambda}
    N_j \bs{\lambda}_j = \mb{n}_j\,.
\end{align}
On the one hand, by Cramer's rule
\begin{align*}
    \lambda_{0,j}=(-1)^{j+1} \frac{\det(N_0)}{\det(N_j)}\,.
\end{align*}
On the other hand, multiplying (\ref{lambda}) by $\mb{v}_j$ gives 
\begin{align*}
    a \lambda_{0,j} = \mb{n}_j \cdot \mb{v}_j\,.
\end{align*}
Since $a > 0$
\begin{align*}
    \sgn(\mb{n}_j \cdot \mb{v}_j) = \sgn\left((-1)^{j+1} \frac{\det(N_0)}{\det(N_j)}\right)
\end{align*}
and thus the relations $R_j$, $j \in [n]$, are fixed.

Using the Hessian normal form we obtain that the distance between $\mb{v}_j$ and the facet not containing $\mb{v}_j$
is $|\mb{n}_j \cdot \mb{v}_j|$ if $ j \in [n]$ and $a$ if $j=0$. Consequently
\begin{align}\label{vol}
    \frac{1}{(n-1)!} |\det(V_0)|=n \vol(\cP)=a A_0=|\mb{n}_1 \cdot \mb{v}_1| A_1=\dots=|\mb{n}_n \cdot \mb{v}_n| A_n\,.
\end{align}
By (\ref{nv}) the matrix $N_0\tp V_0$ is the diagonal matrix with the entries $\mb{n}_j \cdot \mb{v}_j$ in its diagonal, $j \in [n]$. Hence
\begin{align}\label{det}
    |\det(N_0)| |\det(V_0)| = \prod_{j=1}^n |\mb{n}_j \cdot \mb{v}_j|\,.
\end{align}
From (\ref{vol}) and (\ref{det}) we derive
\begin{align*}
    \left(\frac{1}{(n-1)!}\right)^n |\det(V_0)|^n&=a^n A_0^n\\
    & = \prod_{j=1}^n |\mb{n}_j \cdot \mb{v}_j| \prod_{j=1}^n A_j \\
    & = |\det(N_0)| |\det(V_0)| \prod_{j=1}^n A_j \,,
\end{align*}
i.e.,
\begin{align*}
    \frac{1}{(n-1)!} |\det(V_0)|=\sqrt[n-1]{(n-1)!|\det(N_0)| \prod_{j=1}^n A_j}
\end{align*}
and again with (\ref{vol}) we obtain
\begin{align*}
    a=\frac{1}{A_0}\sqrt[n-1]{(n-1)!|\det(N_0)| \prod_{j=1}^n A_j}\,.
\end{align*}
\qed

\begin{figure}[h]
    \centering
    \includegraphics[height=4cm]{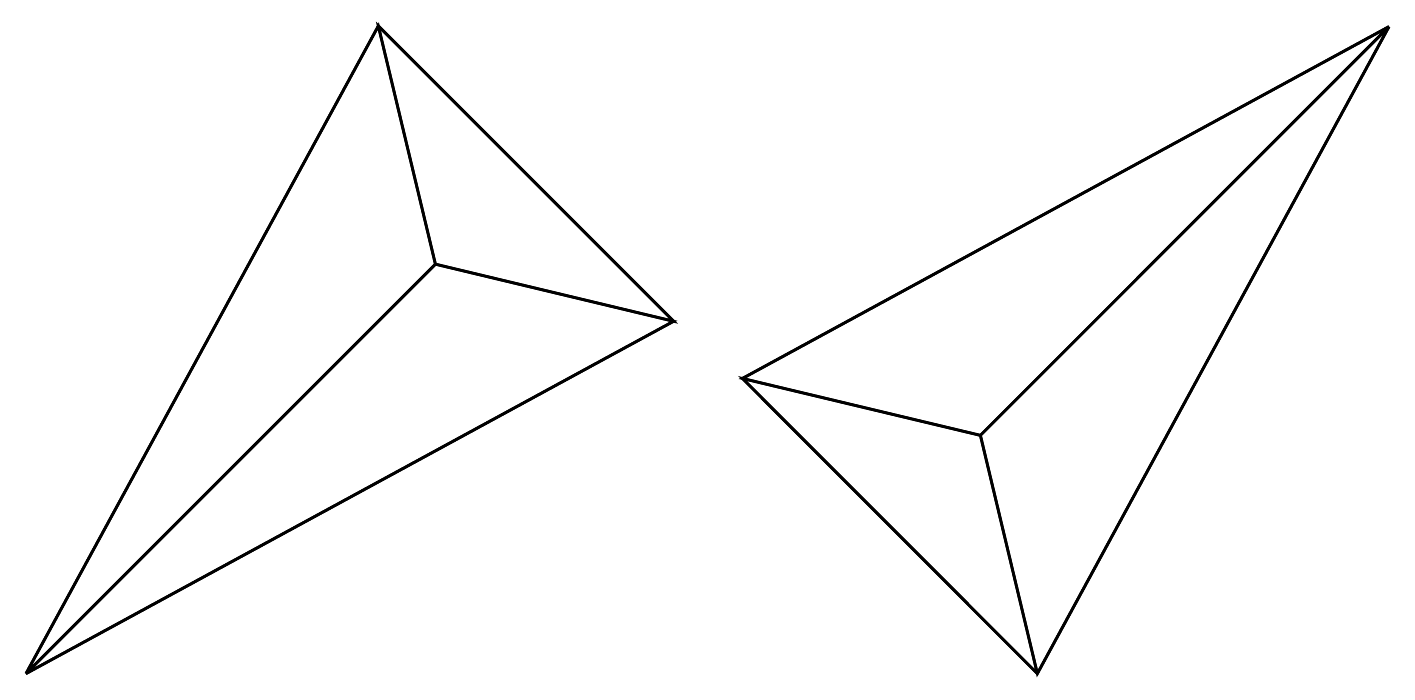}
    \caption{Reconstruction of a $3$-dimensional tetrahedron (Left: original simplex, Right: reconstructed simplex being translated and reflected) based on the modulus of its Fourier transform and the method provided by the proof of Theorem \ref{simplex_reconstruction_theorem}.}
    \label{simplex_3D_explicit}
\end{figure}

\textbf{Example} The proof of Theorem \ref{simplex_reconstruction_theorem} provides a reconstruction method for an unknown simplex given its facet-indicator set $I=\{(\mb{n}_j,A_j): j \in [f]\}$. We approximately computed $I$ using the Fourier transform as in Section \ref{significant_information_section}. The function $\bs{\sigma}: \bR^2 \rightarrow \bR^3$ was chosen as
\begin{align*}
    \bs{\sigma}(\mb{t})= \begin{pmatrix} \sin(t_1)\cos(t_2)\\ \sin(t_1)\sin(t_2)\\ \cos(t_1)\\ \end{pmatrix}, t_1, t_2\in[0, \pi)\,.
\end{align*}
Therefore, the set 
\begin{align*}
    \cS=\{\bs{\sigma}(\mb{t}): \mb{t} \in [0, \pi)\times [0, \pi)\}\,,
\end{align*}
corresponds to a hemisphere. The wave length parameter $\lambda$ was set to $0.01$ again. A comparison between the original and the (translated and reflected) reconstructed terahedron is given by Figure \ref{simplex_3D_explicit}.

For the reconstruction of a more complex facet-generic convex polytope $\cP$ given its facet-indicator set $I$ we need to know its outer normal vectors, respectively its inner normal vectors. Hence, the question arises for which sign variations the vectors $\mb{n}_j$ included by $I$ are only in outward direction. A solution is given by Minkowski's Theorem (see \cite[Section 7.1]{Alexandrov2005}) and Proposition 1 of \cite{Klain2004} formulated as follows (see \cite[Theorem 2]{Klain2004}):
\begin{theorem}\label{minkowski_theorem}
Suppose that $\mb{n}_1, \hdots,\mb{n}_f\in\bR^n$ are unit vectors spanning $\bR^n$ and that $A_1, \hdots, A_f > 0$. Then there exists a closed convex polytope whose facets have outward unit normal vectors $\mb{n}_j$ and corresponding facet areas $A_j$, if and only if
\begin{align}\label{minkowski_condition}
    A_1\mb{n}_1+\dots+A_f\mb{n}_f=\mb{0}.
\end{align}
Moreover, this polytope is unique up to translation.
\end{theorem}

Therefore, given a facet-indicator set $I$ for an unknown convex polytope $\cP$ we have to check all $2^{f-1}$ sign variations for the vectors $\mb{n}_j, j \in [f]$, (inverted variations of already investigated variations do not have to be considered) to extract the outward/inward normal vectors of $\cP$ using condition (\ref{minkowski_condition}). Please note that more than one of the $2^{f-1}$ variations can fulfil condition (\ref{minkowski_condition}) leading to ambiguities as we could see in Figure \ref{2D_example_ambiguity}.

Theorem \ref{minkowski_theorem} states that convex polytopes are fully determined by their outward normal vectors and their facet areas. However, this theorem is not constructive. Hence, the reconstruction requires further considerations. 

In the literature, the term \emph{Extended Gaussian Image (EGI)} is often used for our purposes (see \cite{Horn1984}). The EGI of a convex polytope $\cP$ can be interpreted as a set of vectors including the orientations of the facets, i.e., the outer normal vectors of $\cP$. Furthermore, the length of each vector equals the area of the corresponding facet. Therefore, the EGI of $\cP$ is a set
\begin{align}\label{EGI_set}
    E = \{\mb{m}_j=A_j\mb{n}_j:j\in[f]\}\,,
\end{align}
i.e., having the EGI of $\cP$ is the same like having its facet-indicator set with only outwardly directed normal vectors.

\begin{figure}[h]
    \centering
    \includegraphics[height=5.5cm]{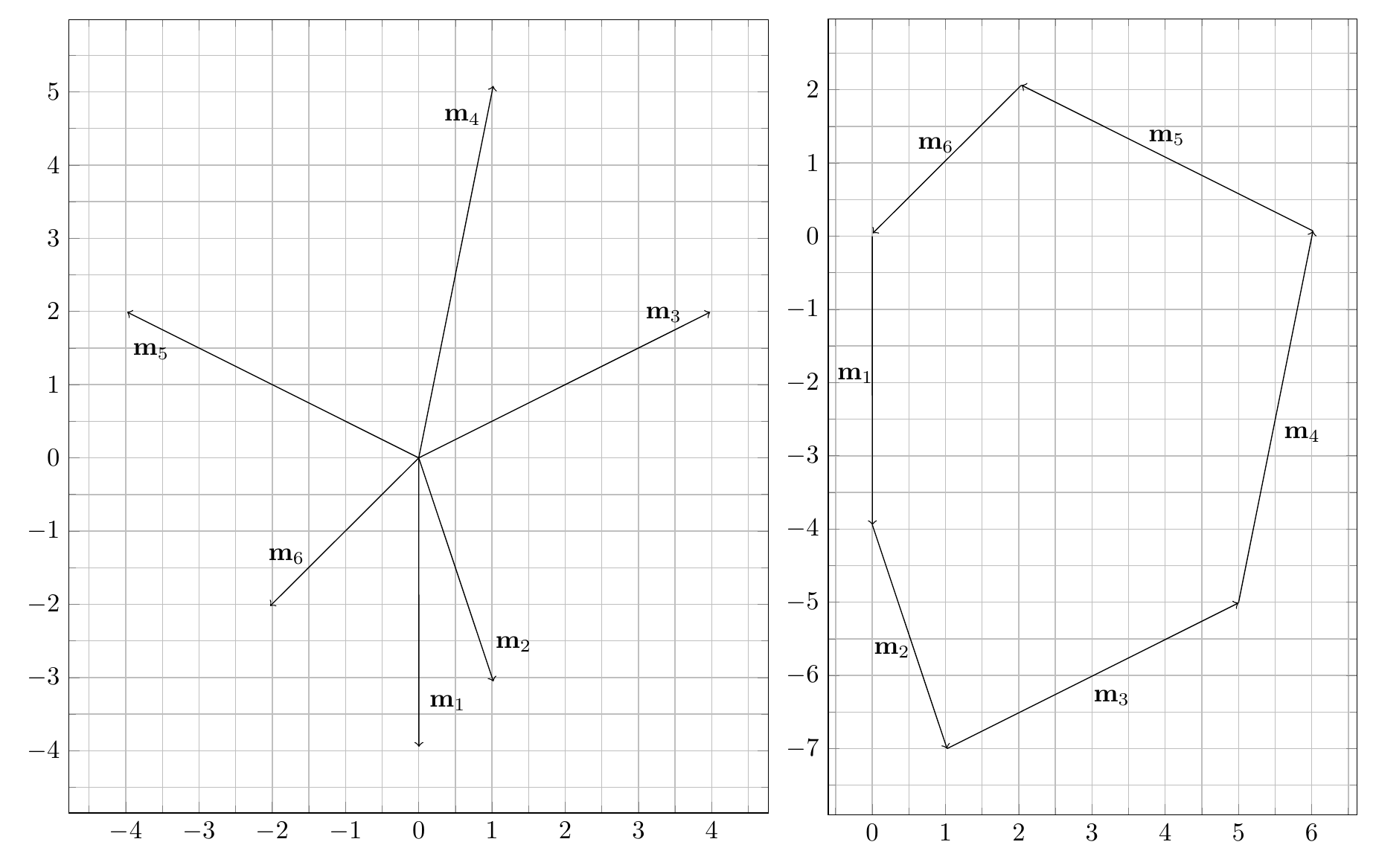}
    \caption{Reconstruction of the $2$-dimensional facet-generic convex polygon given in Figure \ref{2D_example_for_psi} (Left: the Extended Gaussian Image of the polygon with anticlockwise ordered vectors, Right: reconstructed polygon) based on the modulus of its Fourier transform and the method from \cite{Little1983}. Note that the facet-indicator set allows a further reconstruction due to the mentioned ambiguity.}
    \label{polygon_2D_normal}
\end{figure}

For a 2-dimensional unknown facet-generic convex polygon $\cP$ there is a simple reconstruction method given its EGI (see \cite{Little1983} and Figure \ref{polygon_2D_normal}). Assume that the vectors $\mb{m}_j$ are in anticlockwise order. Take $\mb{m}_1$ and place its tail at the origin. Afterwards, take vector $\mb{m}_j$ and place its tail at the head of $\mb{m}_{j-1}$. Because of (\ref{minkowski_condition}) the system sums to zero and we get a closed polygon. At the end rotate the polygon by $\frac{\pi}{2}$. Due to the construction of the set (\ref{EGI_set}) the polygon will have the correct orientation and the edges the correct lengths.

The extension of the $2$-dimensional algorithm to higher dimensions is not possible since the adjacencies between the facets are in higher dimensions not trivially given by the EGI. However, for a 3-dimensional facet-generic convex polytope there are also existing reconstruction methods using the outward normal vectors and the corresponding facet areas of an unknown polytope.

\begin{figure}[h]
    \centering
    \includegraphics[height=4cm]{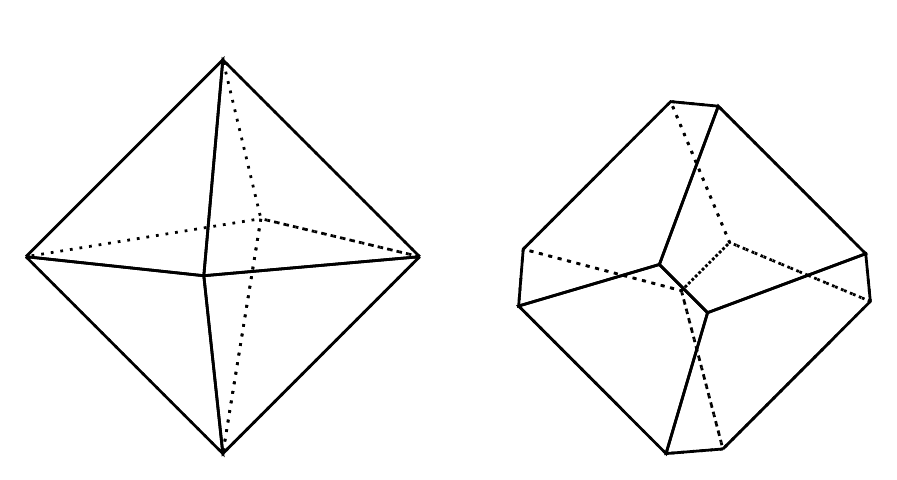}
    \caption{Reconstruction of a $3$-dimensional facet-generic convex polytope (Left: original polytope, Right: reconstructed polytope) based on the modulus of its Fourier transform and the method from \cite{Sellaroli2017}. The original polytope is a slightly deformed octahedron to avoid parallel facets.}
    \label{polytope_3D_sellaroli}
\end{figure}

For example, the algorithms in \cite{Ikeuchi1981}, \cite{Little1983} and \cite{Moni1990} are based on the EGI. Moreover, in \cite{Alexandrov2005b} another reconstruction algorithm was developed using Blaschke sums. The method in \cite{Sellaroli2017} is based on a method by Lasserre to compute the volume of a convex polytope. Please note, that the cited papers assume the exact vectors and areas but in our scenario the normal vectors and facet areas are only approximations. Hence, there is a lack of robustness (see Figure \ref{polytope_3D_sellaroli}, where the reconstruction algorithm from \cite{Sellaroli2017} was used).

\section{Conclusion}

This paper investigates how to reconstruct significant properties of an unknown facet-generic convex polytope $\cP$ given the modulus of its Fourier transform with small wave length on a complete subset $\cS$ of $\bR^n$. It turns out that it is possible to compute approximately a normal vector of each facet of $\cP$ and the corresponding facet area. Furthermore, if $\cP$ is an $n$-dimensional simplex a unique reconstruction is possible up to translation and reflection in a point. Since with this approach one cannot distinguish between inward and outward directions of the normal vectors, uniqueness is not guaranteed for arbitrary $n$-dimensional polytopes. Finally, existing reconstruction algorithms for given outward normal vectors and facet areas in the $2$- and $3$-dimensional case are briefly reported and applied.

It remains open how to reconstruct (significant properties of) convex polytopes with parallel facets, i.e., not facet-generic polytopes. 
Moreover an estimation of the approximation error as well as a robustness analysis are challenging problems for the future.

It should be noted that in actual experiments one has only approximate values of the modulus of the Fourier transform an the Ewald \emph{half-}sphere, i.e., a half-sphere touching a coordinate hyperplane at the origin, where in addition a neighborhood of the origin is deleted. Hence not all potential normal vectors can be represented by this part of the sphere, i.e., there is a lack of information. 
It is an interesting physical problem to extend the experiments in such a way, such the necessary values can be obtained on a larger part of the Ewald sphere.

\section*{Acknowledgement}

This work was partly supported by the European Social Fund (ESF) and the Ministry of Education, Science and Culture of Mecklenburg-Western Pomerania (Germany) within the project NEISS – Neural Extraction of Information, Structure and Symmetry in Images under grant no ESF/14-BM-A55-0006/19.

\bibliography{references}

\begin{thebibliography}{10}

\bibitem{Alexandrov2005}
Alexander~D. Alexandrov.
\newblock {\em Convex Polyhedra}.
\newblock Springer Monographs in Mathematics. Springer, Berlin, Heidelberg,
  2005.

\bibitem{Alexandrov2005b}
Victor Alexandrov, Natalia Kopteva, and Semen~S. Kutateladze.
\newblock Blaschke addition and convex polyhedra.
\newblock ArXiv:math/0502345, 2005.

\bibitem{Barke2015}
Ingo Barke, Hannes Hartmann, Daniela Rupp, Leonie Fl{\"u}ckiger, Mario Sauppe,
  Marcus Adolph, Sebastian Schorb, Christoph Bostedt, Rolf Treusch, Christian
  Peltz, Stephan Bartling, Thomas Fennel, Karl-Heinz Meiwes-Broer, and Thomas
  M\"oller.
\newblock The 3{D}-architecture of individual free silver nanoparticles
  captured by {X}-ray scattering.
\newblock {\em Nature communications}, 6(1):1--7, 2015.

\bibitem{Barvinok2008}
A.~Barvinok.
\newblock {\em Integer points in polyhedra}.
\newblock European Math. Soc. Publ. House, Z\"{u}rich, 2008.

\bibitem{Beck2007}
M.~Beck and S.~Robins.
\newblock {\em Computing the continuous discretely}.
\newblock Springer, Berlin et al, 2009.

\bibitem{Beinert2017}
Robert Beinert and Gerlind Plonka.
\newblock Sparse phase retrieval of one-dimensional signals by {P}rony's
  method.
\newblock {\em Frontiers in Applied Mathematics and Statistics}, 3:5, 2017.

\bibitem{Engel2020}
Konrad Engel and Bastian Laasch.
\newblock The modulus of the {F}ourier transform on a sphere determines
  3-dimensional convex polytopes.
\newblock ArXiv:2009.10414, 2020.

\bibitem{Gravin2012}
Nick Gravin, Jean Lasserre, Dmitrii~V. Pasechnik, and Sinai Robins.
\newblock The inverse moment problem for convex polytopes.
\newblock {\em Discrete \& Computational Geometry}, 48(3):596--621, 2012.

\bibitem{Gravin2018}
Nick Gravin, Dmitrii~V. Pasechnik, Boris Shapiro, and Michael Shapiro.
\newblock On moments of a polytope.
\newblock {\em Anal. Math. Phys.}, 8:255--287, 2018.

\bibitem{Horn1984}
Berthold~K.P. Horn.
\newblock Extended {Ga}ussian {I}mages.
\newblock {\em Proceedings of the IEEE}, 72(12):1671--1686, 1984.

\bibitem{Ikeuchi1981}
Katsushi Ikeuchi.
\newblock Recognition of 3-{D} objects using the {E}xtended {G}aussian {I}mage.
\newblock In {\em IJCAI}, pages 595--600, 1981.

\bibitem{Klain2004}
Daniel~A. Klain.
\newblock The {M}inkowski problem for polytopes.
\newblock {\em Advances in Mathematics}, 185(2):270--288, 2004.

\bibitem{Little1983}
James~J. Little.
\newblock An iterative method for reconstructing convex polyhedra from
  {E}xtended {G}aussian {I}mages.
\newblock In {\em Proceedings of the Third AAAI Conference on Artificial
  Intelligence}, pages 247--250, 1983.

\bibitem{Moni1990}
Shankar Moni.
\newblock A closed-form solution for the reconstruction of a convex polyhedron
  from its {E}xtended {G}aussian {I}mage.
\newblock In {\em [1990] Proceedings. 10th International Conference on Pattern
  Recognition}, volume~1, pages 223--226. IEEE, 1990.

\bibitem{Raines2010}
Kevin~S. Raines, Sara Salha, Richard~L. Sandberg, Huaidong Jiang, Jose~A.
  Rodr{\'\i}guez, Benjamin~P. Fahimian, Henry~C. Kapteyn, Jincheng Du, and
  Jianwei Miao.
\newblock Three-dimensional structure determination from a single view.
\newblock {\em Nature}, 463(7278):214--217, 2010.

\bibitem{Rossbach2019}
J\"org Rossbach, Jochen~R. Schneider, and Wilfried Wurth.
\newblock 10 years of pioneering {X}-ray science at the free-electron laser
  {FLASH} at {DESY}.
\newblock {\em Physics Reports}, 808:1--74, 2019.

\bibitem{Seibert2011}
M.~Marvin Seibert, Tomas Ekeberg, Filipe~RNC Maia, Martin Svenda, Jakob
  Andreasson, Olof J{\"o}nsson, Du{\v{s}}ko Odi{\'c}, Bianca Iwan, Andrea
  Rocker, Daniel Westphal, et~al.
\newblock Single mimivirus particles intercepted and imaged with an {X}-ray
  laser.
\newblock {\em Nature}, 470(7332):78--81, 2011.

\bibitem{Sellaroli2017}
Giuseppe Sellaroli.
\newblock An algorithm to reconstruct convex polyhedra from their face normals
  and areas.
\newblock ArXiv:1712.00825, 2017.

\bibitem{Stielow2020}
Thomas Stielow, Robin Schmidt, Christian Peltz, Thomas Fennel, and Stefan
  Scheel.
\newblock Fast reconstruction of single-shot wide-angle diffraction images
  through deep learning.
\newblock {\em Machine Learning: Science and Technology}, 1:045007, 2020.

\bibitem{Wischerhoff2016}
Marius Wischerhoff and Gerlind Plonka.
\newblock Reconstruction of polygonal shapes from sparse {F}ourier samples.
\newblock {\em Journal of Computational and Applied Mathematics}, 297:117--131,
  2016.

\bibitem{Wuttke2017}
Joachim Wuttke.
\newblock Form factor ({F}ourier shape transform) of polygon and polyhedron.
\newblock ArXiv:1703.00255, 2017.

\end{thebibliography}
\bibliographystyle{plain}

\end{document}